\documentclass[prl,amsmath,amssymb,superscriptaddress,twocolumn]{revtex4-1}
\usepackage{soul}
\usepackage{graphicx}
\usepackage{epsfig}
\usepackage{dcolumn}
\usepackage{bm}
\usepackage{rotating}
\usepackage[table]{xcolor}
\usepackage[english]{babel}
\begin{document}
\widetext 
\title{Non-adiabatic Kohn Anomaly in Heavily Boron-doped Diamond}
\author{Fabio Caruso}
\affiliation{Department of Materials, University of Oxford, Parks Road, Oxford, OX1 3PH, United Kingdom}
\author{Moritz Hoesch}
\affiliation{Diamond Light Source, Harwell Campus, Didcot OX11 0DE, United Kingdom}
\author{Philipp Achatz}
\affiliation{Univ. Grenoble Alpes, CNRS, Inst. NEEL, F-38000 Grenoble, France}
\author{Jorge Serrano}
\affiliation{Yachay Tech University, School of Physical Sciences and Nanotechnology, 100119-Urcuqu\'{i}, Ecuador}
\author{Michael Krisch}
\affiliation{European Synchrotron Radiation Facility, 6 rue Jules Horowitz, 38043 Grenoble Cedex, France}
\author{Etienne Bustarret}
\affiliation{Univ. Grenoble Alpes, CNRS, Inst. NEEL, F-38000 Grenoble, France}
\author{Feliciano Giustino}
\email{{feliciano.giustino@materials.ox.ac.uk}}
\affiliation{Department of Materials, University of Oxford, Parks Road, Oxford, OX1 3PH, United Kingdom}
\date{\today}

\date{\today}
\begin{abstract}
We report evidence of a non-adiabatic Kohn anomaly in boron-doped diamond,
using a joint theoretical and experimental analysis of the phonon dispersion
relations. We demonstrate that standard calculations of phonons using density
functional perturbation theory are unable to reproduce the 
dispersion relations of the high-energy phonons measured by high-resolution
inelastic x-ray scattering. On the contrary, by taking into account non-adiabatic
effects within a many-body field-theoretic framework, we obtain excellent agreement
with our experimental data. 
This result indicates a 
breakdown of the Born-Oppenheimer approximation in the 
phonon dispersion relations of boron-doped diamond. 
\end{abstract}
\maketitle

The Kohn anomaly (KA) is one of the most striking manifestations of the influence 
of electron-phonon coupling on the lattice dynamics of metals \cite{Kohn1959}. 
KAs result from the screening of lattice vibrations by 
virtual electronic excitations across the Fermi surface \cite{Mahan2000},
and manifest themselves through distinctive dips in the phonon dispersion relations.
The existence of KAs was confirmed by inelastic neutron scattering experiments 
\cite{Brockhouse1961} shortly after Kohn's theoretical prediction \cite{Kohn1959}. 
Since then KAs have been observed in a number of metals \cite{Brockhouse1962,Nakagawa1963,Koenig1964},
conventional superconductors \cite{Baron2004,Aynajian1509}, as well as 
superconducting semiconductors \cite{Hoesch2007}.

Interest in KAs was recently
re-ignited by the discovery of {\it non-adiabatic} KAs in carbon materials,
such as graphene \cite{Lazzeri2006,Pisana2007}, carbon nanotubes \cite{Caudal2007,Piscanec2007},
and graphite intercalation compounds \cite{Calandra2007,Saitta2008,Calandra2010}. 
At variance with adiabatic KAs, which are 
well described in the adiabatic Born-Oppenheimer approximation \cite{Kohn1959},  
non-adiabatic KAs arise when the electronic screening takes place on timescales
which are comparable to the period of lattice vibrations, and signal 
the breakdown of the Born-Oppenheimer approximation. In the majority of
current first-principles calculations, these non-adiabatic effects are ignored
on the grounds that they should be of the order of $m/M$, with $m$ the electron
mass and $M$ the characteristic nuclear mass.
{While the calculations of non-adiabatic 
phonon linewidths may be performed using standard implementations \cite{giustino2016}, 
first-principles studies of renormalization effects on the phonon dispersions 
due to non-adiabaticity are extremely challenging, and 
have thus far been confined to low-dimensional compounds.
In particular, for metallic compounds characterized by a two-dimensional, quasi-two-dimensional,
or one-dimensional structure it has been shown that non-adiabatic effects can
alter significantly the phonon dispersion relations 
\cite{Lazzeri2006,Pisana2007,Caudal2007,Piscanec2007,Calandra2007,Saitta2008,Calandra2010,Leroux2015}. 
}
Instead, for three-dimensional bulk metals, 
it has been suggested that non-adiabatic effects might be too small to be 
observable in experiment \cite{Saitta2008}. 

{
The strong coupling between electrons and longitudinal optical (LO) phonons 
in diamond, manifested for instance by a 0.6~eV zero-point motion 
band-gap renormalization \cite{Giustino2010prl,Marini2011prl,Antonius2015prl} and the emergence of type-II superconductivity 
for sufficiently high B-doping \cite{Ekimov2004nature}, make it 
a good candidate for the observation of non-adiabatic effects in the 
phonon dispersions. 
Pristine diamond has previously attracted considerable interest 
due to the anomalous overbending of the optical phonon branch \cite{Schwoerer1998}. 
In presence of B-dopants, 
}
the electron-phonon interaction induces a softening of the 
LO phonons at long wavelengths, and a concomitant broadening 
of the spectral lines \cite{Hoesch2007,Bustarret2015}. 
These effects are taken to be the signatures of a doping-induced KA. 
The measured softening is found to be 
between 4 and 7~meV for B-doping concentrations of 
$10^{20}$-$10^{21}$~cm$^{-3}$ \cite{Hoesch2007,Bustarret2015}.
{Intriguingly, first-principles
calculations \cite{Boeri2004prl,Blase2004prl,Pickett2004prl,Ma2005,GiustinoPRL2007} 
gave considerably more pronounced phonon softening, in the
range of 20 to 30~meV.
This unusually large discrepancy between experiment and theory
remains an outstanding question in the physics of superconducting diamond \cite{sacepe2006}.
This led us to formulate the hypothesis that in order to explain the measured KA in diamond 
it might be necessary to invoke non-adiabatic effects.

In this work we analyze the dispersion relations of 
the longitudinal-optical (LO) phonons of B-doped diamond using 
state-of-the-art first-principles calculations and inelastic x-ray scattering (IXS) measurements.
By comparing theory and experiment we demonstrate that the non-adiabatic correction 
to the LO phonon energy is indeed very large, up to 10~meV. After including non-adiabatic effects
within a field-theoretic framework, we obtain an unprecedented agreement 
between theory and experiment, and we resolve 
the discrepancy between earlier theoretical works and measured phonon dispersions.
{Our results 
demonstrate a breakdown of the adiabatic Born-Oppenheimer approximation 
in the {phonon dispersion relations} of boron-doped diamond, revealing that 
these effects may be sizeable also in three-dimensional bulk compounds.} 
}

The B-doped diamond samples were prepared by microwave plasma-enhanced chemical vapor deposition (MPCVD) 
from a hydrogen-rich gas phase with added diboran ($\mathrm{B}_{2}\mathrm{H}_{6}$). The samples were 
grown homoepitaxially on type Ib synthetic crystals with (001) oriented surfaces at thicknesses 
of $25\pm5 \ \mu$m \cite{Achatz2010814}. The boron concentration was determined from secondary ion mass spectroscopy 
(SIMS) of $^{11}\mathrm{B}^{-}$, $^{12}\mathrm{C}^{-}$ and $^{11}\mathrm{B}^{12}\mathrm{C}^{-}$ ions.
{For a B-doping concentration of $1.4\cdot10^{21}$~cm$^{-3}$, the samples exhibit superconducting
behaviour with critical temperature $T_c = 2.8$~K}.
IXS spectra were measured at beamline ID28 at the European Synchrotron 
Radiation Facility (ESRF) with an energy resolution of 3.2 meV. The samples were aligned {with the beam 
directed parallel to the surface and passing through the substrate} or the B-doped diamond film, for measurements of pristine diamond and 
B-doped diamond, respectively. The scattering vector ${\bf Q}$ was varied from 
{$(2.06,0,0)2\pi/a$ (close to $\Gamma$) to $(3,-0.12,0)2\pi/a$
(close to $X$), with $a=3.67$~\AA. The small deviations in the $(0,k,0)$ direction 
are given in Supplemental Table~1 \cite{sup}.} 
The measured IXS spectra are shown in Fig.~\ref{fig1}~(c)-(e) as heat maps, {and in
Supplemental Fig.~1 as individual scans} \cite{sup}. 
For the undoped case, our measurements are in excellent agreement with previous experimental data \cite{Kulda2002}.

\begin{figure*}
\begin{center}
\includegraphics[width=0.98\textwidth]{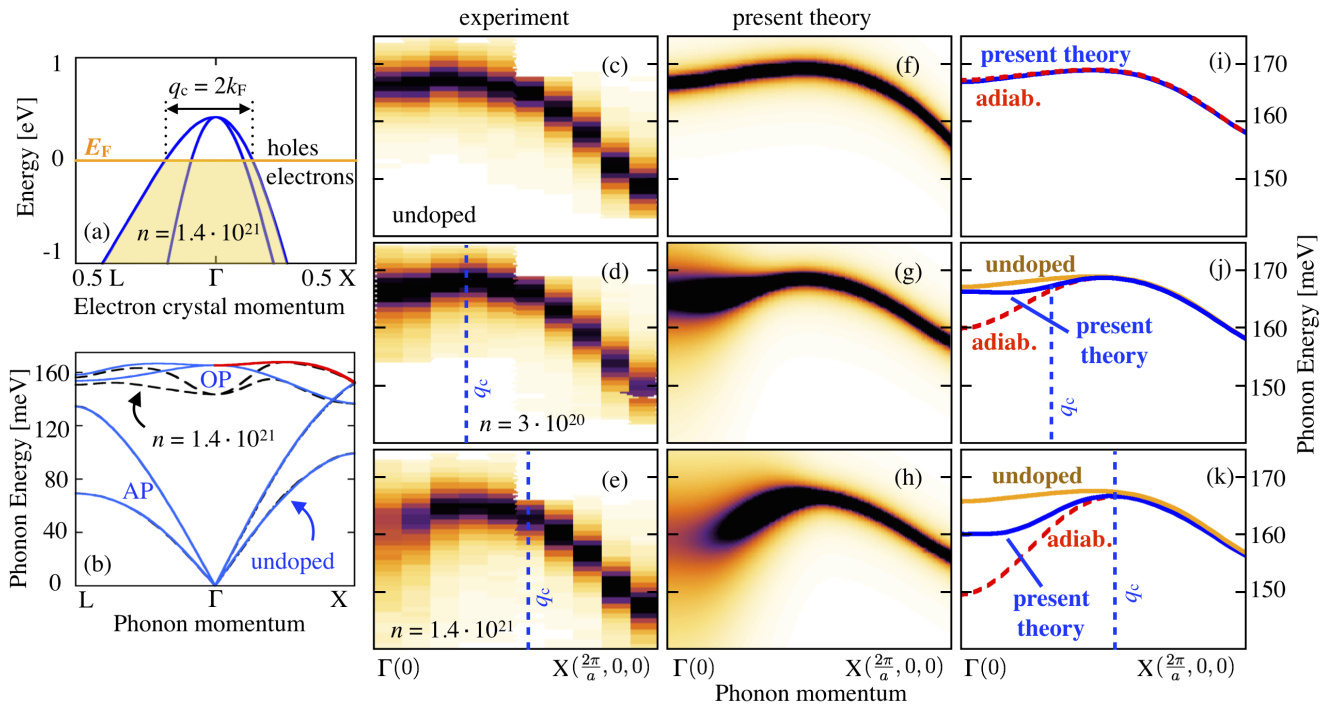}
\caption{\label{fig1} 
(a)~Density-functional theory 
band structure of diamond for a B concentration of $1.4\cdot10^{21}$~cm$^{-3}$. 
(b)~{Adiabatic} phonon dispersions of pristine 
(blue lines) and B-doped diamond (dashed black lines) 
for momenta along L-$\Gamma$-X, as obtained from density-functional perturbation theory. 
(c)-(e)~Measured IXS spectra of pristine and B-doped diamond. 
The critical momentum for the onset of the KA, $q_c=2k_{\rm F}$, 
is indicated by vertical dashed lines, see also (a). 
(f)-(h)~Non-adiabatic spectral function, obtained from Eqs.~(\ref{eq:PiNA})-(\ref{eq:phA}), 
for the LO phonon of (c) pristine and (d)-(e) B-doped diamond along $\Gamma$-X. 
The phonon branch considered here is marked by the red line in panel (b). 
(i)-(k)~Phonon energies obtained from Eq.~(\ref{eq:NAph}) in the 
adiabatic approximation ($ \Pi^{\rm NA}_{{\bf q}\nu}=0$), and 
from the fully non-adiabatic theory (present theory). 
Non-adiabatic phonon dispersions of undoped diamond are reported for comparison. 
All doping concentrations are in units of cm$^{-3}$. 
}
\end{center}
\end{figure*}

Non-adiabatic phonon dispersions were computed 
from first-principles within the many-body theory of 
electron-phonon coupling. Non-adiabatic effects were accounted for  
via the phonon self-energy $\Pi_{{\bf q}\nu}^{\rm NA}$ \cite{giustino2016}: 
\begin{align}\label{eq:PiNA}
\hbar\Pi_{{\bf q}\nu}^{\rm NA} &(\omega) = 
2 \sum_{mn} \int \frac{d{\bf k}}{\Omega_{\rm BZ}} 
g_{mn,\nu}^{\rm b}({\bf k,q}) 
g_{mn,\nu}^{*}({\bf k,q})\\ 
&\times\left[
\frac{ f_{n{\bf k}} - f_{m{\bf k+q}} }{
\epsilon_{m{\bf k+q}} - \epsilon_{n{\bf k}} - \hbar( \omega+i\eta)}
-
\frac{ f_{n{\bf k}} - f_{m{\bf k+q}} }{
\epsilon_{m{\bf k+q}} - \epsilon_{n{\bf k}}}\right],\nonumber
\end{align}
where $\epsilon_{n{\bf k}}$ and $f_{n{\bf k}}$ denote single-particle energies 
and Fermi-Dirac occupation factors, $\eta$ is a positive infinitesimal, 
and $\Omega_{\rm BZ}$ is the Brillouin zone volume.  
The {\it screened} electron-phonon matrix elements $g_{mn,\nu}({\bf k,q})$ were obtained as 
$g_{mn,\nu}({\bf k,q}) = ( {\hbar}/{2M \omega_{{\bf q}\nu} })^{1/2} 
 \langle \psi_{m{\bf k+q}}|\partial_{{\bf q}\nu}V | \psi_{n{\bf k}}\rangle$, 
where $\psi_{n{\bf k}}$ denote Kohn-Sham single-particle eigenstates, $M$ the C mass,  
and $\partial_{{\bf q}\nu}V$ the derivative of the self-consistent 
potential associated with the $\nu$-th phonon mode 
with wavevector ${\bf q}$ and energy $\hbar\omega_{{\bf q}\nu}$.
$g_{mn,\nu}({\bf k,q})$ is obtained from the
{\it bare} matrix element $g^{\rm b}_{mn,\nu}({\bf k,q})$ by screening the variation of the
ionic potential using the electronic dielectric function. 
Here we calculate $g^{\rm b}_{mn,\nu}({\bf k,q})$
by unscreening $g_{mn,\nu}({\bf k,q})$  and neglect local-field effects for simplicity.
Equation~(\ref{eq:PiNA}) accounts for both the {\it screened} and the {\it bare} 
electron-phonon vertices ($g$ and $g^{\rm b}$) and it thus avoids the 
approximation employed in previous first-principles {calculations}, 
whereby the matrix elements  $g_{mn,\nu}^{\rm b}({\bf k,q}) g_{mn,\nu}^{*}({\bf k,q})$ 
were replaced by $|g_{mn,\nu}({\bf k,q})|^2$ \cite{giustino2016}.
The non-adiabatic phonon dispersions, that is, the dispersions modified by the phonon self-energy of Eq.~(\ref{eq:PiNA}), 
were extracted directly from the phonon spectral function 
\footnote{Calculations were performed using density-functional theory \cite{hohenbergkohn,kohnsham1965} within the Perdew-Burke-Ernzerhof 
generalized-gradient approximation \cite{PBE} for the exchange-correlation functional, as implemented in 
{\tt Quantum Espresso} \cite{Giannozzi2009}. 
We used a plane-wave basis set with a kinetic energy cutoff of 60~Ry, 
norm-conserving Goedecker-Hartwigsen-Hutter-Teter pseudopotentials \cite{Hartwigsen1998}, and 
a 8$\times$8$\times$8 Monkhorst-Pack grid for sampling the Brillouin zone.
Adiabatic phonon frequencies and eigenvectors were computed through 
density-functional perturbation theory \cite{Baroni2001} 
on a 6$\times$6$\times$6 grid. 
Electron bands, phonon dispersions, and electron-phonon matrix elements were
interpolated using maximally localized Wannier functions \cite{Marzari2012,Mostofi2008}, 
and Eq.~(\ref{eq:PiNA}) 
was computed using {\tt EPW} v4 \cite{Giustino2007prb,ponce2016}. 
The Brillouin-zone summation in Eq.~(\ref{eq:PiNA}) was evaluated 
using one million random ${\bf k}$-points, and a broadening parameter 
$\hbar\eta=10$~meV.
In all calculations, doping with boron was modelled in the 
rigid-band approximation through a shift of the Fermi level below the 
valence band top. A temperature of 300~K was included via the 
Fermi-Dirac occupation factors in Eq.~(\ref{eq:PiNA}).
To approximately account for finite energy and momentum resolution, 
the results of Eqs.~(\ref{eq:phA})-(\ref{eq:NAph}) were broadened  
by  $\Delta E= 1~$meV and $\Delta k = 0.08$~\AA$^{-1}$ via 
a Gaussian convolution.
}:
\begin{align}\label{eq:phA}
A_{{\bf q}\nu} (\omega) = \pi^{-1} {\rm Im\,}\left[\frac{2\omega_{{\bf q}\nu}}
{ \omega^2 - \omega^2_{{\bf q}\nu} - 2 \omega_{{\bf q}\nu}  \Pi^{\rm NA}_{{\bf q}\nu}(\omega)} \right]. 
\end{align}
Equation~(\ref{eq:phA}), which constitutes the phonon counterpart of the electronic spectral function \cite{Mahan2000}, 
exhibits peaks at the non-adiabatic phonon frequencies $\Omega_{{\bf q}\nu}$ given by:
\begin{align}\label{eq:NAph}
\Omega_{{\bf q}\nu}^2  \simeq \omega^2_{{\bf q}\nu} + 2 \omega_{{\bf q}\nu} 
{\rm Re\,}\Pi^{\rm NA}_{{\bf q}\nu}(\Omega_{{\bf q}\nu}),
\end{align}
with a full-width at half-maximum $\Gamma_{{\bf q}\nu} =2 \hbar\, {\rm Im} \Pi^{\rm NA}_{{\bf q}\nu}(\Omega_{{\bf q}\nu})$.
Non-adiabatic phonon spectral functions obtained from Eq.~(\ref{eq:phA}) are reported in Fig.~\ref{fig1}~(f)-(h), 
whereas the phonon dispersions derived from Eq.~(\ref{eq:NAph}) are shown in Fig.~\ref{fig1}~(i)-(k).

Inspection of Eq.~(\ref{eq:PiNA}) reveals that non-adiabatic effects may 
become important whenever the transition energies between occupied and empty 
electronic states ($\epsilon_{m{\bf k+q}} - \epsilon_{n{\bf k}}$) 
approach the characteristic phonon energy $\hbar \omega_{{\bf q}\nu}$.
As in solids $\hbar \omega_{{\bf q}\nu}$ is typically $\lesssim 100$~meV, 
this condition is only satisfied in metals, doped semiconductors,  and 
narrow-gap semiconductors, whereby low-energy 
intra-band transitions may be excited.  
Therefore, in these systems one may expect to observe 
(i) phonon damping effects, with a characteristic timescale set by 
the phonon lifetime $\tau_{{\bf q}\nu}=\hbar/\Gamma_{{\bf q}\nu}$;
and (ii) a renormalization of the adiabatic phonon frequencies, arising 
from the finite value of  ${\rm Re} \Pi^{\rm NA}_{{\bf q}\nu}(\Omega_{{\bf q}\nu})$ in Eq.~(\ref{eq:NAph}). 
On the other hand, the standard Born-Oppenheimer approximation is recovered in the limit $\Pi^{\rm NA}_{{\bf q}\nu} =0$.

Calculations were performed using density-functional
theory (ground state and band structures) and density-functional
perturbation theory (phonon dispersion relations and electron-phonon
matrix elmenents), using Quantum Espresso \cite{Giannozzi2009}, EPW \cite{ponce2016}, and Wannier90 \cite{Mostofi2008}.
The doping was modelled in the rigid-band approximation, and the
spectral functions were computed at 300~K. Complete
calculation details are given in Ref.~[34].
The phonon dispersions of pristine diamond in the adiabatic approximation 
are presented in Fig.~\ref{fig1}~(b) for momenta along the L-$\Gamma$-X path. 
{The  acoustic and optical phonon branches, which correspond 
to the in- and out-of-phase oscillation of the diamond sublattices, are denoted
as AP and OP in Fig.~\ref{fig1}~(b). 
}
Pristine diamond is an insulator with a fundamental band 
gap  $E_{g} =5.4$~eV \cite{Clark312,Zollner1992} and
{ the large optical phonon energy of $\hbar\omega_{\rm ph}=164$~meV reflects
the stiffness of its covalent bonds}. 
Since $E_{g}\gg \hbar \omega_{\rm ph}$, 
non-adiabatic effects are relatively unimportant, and 
the non-adiabatic corrections are smaller than 0.4~meV, {see Fig.~\ref{fig1}~(i)}. 
The resulting phonon dispersions are in excellent agreement 
with our measured IXS spectrum in Fig.~\ref{fig1}~(c), 
in line with the notion that phonons in wide band-gap insulators are well 
described in the adiabatic approximation. 

To quantify the importance of non-adiabaticity for undoped
semiconductors and insulators, we derive a simple estimate of 
the energy renormalization. 
In the limit of non-dispersive electronic bands, 
one may replace $\epsilon_{m{\bf k+q}} - \epsilon_{n{\bf k}} = E_g$ in Eq.~(\ref{eq:PiNA}).
If we further assume an Einstein model for the optical phonons $\hbar\omega_{{\bf q}\nu}=\hbar\omega_{\rm E}$ 
and we restrict ourselves to the limit $\hbar\omega_{\rm E}\ll E_g$,
the term in squared bracket in Eq.~(\ref{eq:PiNA}) reduces to 
$\hbar\omega_{\rm E}/E_g^2$ to first order. 
An explicit approximation for Eq.~(\ref{eq:PiNA}) then is promptly obtained:
${\hbar \Pi} = 2 \epsilon_\infty g^2 {\hbar\omega_{\rm E}}/{E_g^2},$
with $\epsilon_\infty$ being the dielectric constant and $g$ the average electron-phonon matrix element.
For diamond, using $\epsilon_\infty= 5.44$, $E_g= 5.4$~eV, ${\hbar\omega_{\rm E}}=0.16 $~eV, and 
$g= 0.1 $~eV, we obtain $\hbar \Pi = 0.5 $~meV, which is consistent with the first principles calculations 
shown in Fig.~\ref{fig1}~(i).

As compared to the undoped case, the IXS spectra of B-doped diamond 
in Figs.~\ref{fig1}~(d)-(e) exhibit a red-shift 
of the LO phonon energy and an increase of the phonon linewidth close to $\Gamma$, 
which indicate the emergence of a doping-induced KA. 
To quantify the effect of doping on the phonon energy, we define the 
phonon softening parameter $\Delta \Omega_{{\bf q}\nu}(n)=\Omega_{{\bf q}\nu}(0) - \Omega_{{\bf q}\nu}(n)$, 
where $\Omega_{{\bf q}\nu}(n)$ denotes the phonon frequency at a carrier density~$n$. 
The softening and linewidth become more pronounced with the increase of doping concentration.  
The KA is observed only for wave-vectors
smaller than a critical cutoff value
$q_{c}=2k_{\rm F}$, with $k_{\rm F}$ being
the Fermi momentum, which corresponds to the maximum momentum
transfer for electron-phonon scattering on the Fermi surface, see Fig.~\ref{fig1}~(a) \cite{Kohn1959}.
Using the Fermi momentum of the homogeneous electron gas model, 
$k_{\rm F}= (3\pi^2 n/N_m)^{\frac{1}{3}}$, where $N_m=3$ 
is the degeneracy of the valence-band top of diamond, we obtain 
$q_c = 0.3$ and $0.5$~\AA$^{-1}$ for doping levels of 
$3\cdot10^{20}$ and $1.4\cdot10^{21}~{\rm cm}^{-3}$, respectively. 
These values are marked by vertical dashed lines in Fig.~\ref{fig1}~(d)-(e) and (j)-(k).
%

\begin{figure}
\begin{center}
\includegraphics[width=0.48\textwidth]{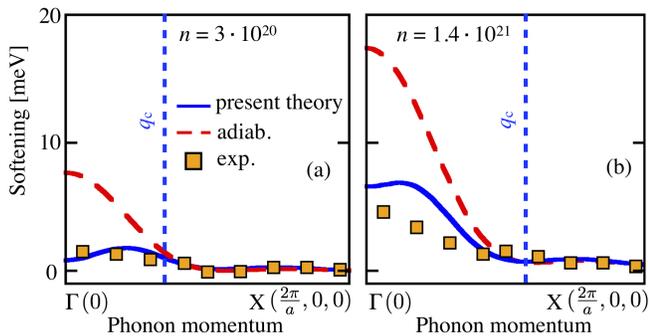}
\caption{\label{fig2} 
Energy renormalization of the longitudinal optical phonons of diamond, 
for doping concentrations of (a) $1.4\cdot10^{21}~{\rm cm}^{-3}$  and 
(b) $3\cdot10^{20}~{\rm cm}^{-3}$: experiment (squares), 
adiabatic (dashed red line) and non-adiabatic theory (blue line). 
}
\end{center}
\end{figure}

For momenta $q < q_c$ 
we find adiabatic phonon dispersions consistent with previous 
works \cite{Boeri2004prl,Ma2005,GiustinoPRL2007}. 
As reported in Refs.~\onlinecite{Hoesch2007,GiustinoPRL2007}, 
however, the adiabatic approximation leads to a systematic 
underestimation of the phonon energy as compared to experiment, which becomes 
more pronounced with the increase of doping concentration. 
Conversely, fully non-adiabatic calculations yield phonon energies in 
excellent agreement with IXS, as revealed by 
the comparison between Fig.~\ref{fig1}~(d)-(e) and (j)-(k). 
To quantify the importance of non-adiabatic effects, we compare in Fig.~\ref{fig2} 
the softening $\Delta \Omega_{{\bf q}\nu}$ and the lineshapes 
for the LO phonon of B-doped diamond, as obtained
from IXS, from the adiabatic approximation, and from 
fully non-adiabatic calculations. 
Above the threshold $q>q_c$ for the onset of the 
KA, theory and experiment yield a
phonon softening smaller than 1~meV for all doping concentrations. 
For $q<q_c$, instead, the positive phonon softening 
reflects the red-shift of the phonon frequency 
induced by electron-phonon interactions. 
Figure~\ref{fig2}~(a)-(b) reveal that the adiabatic approximation 
overestimates the experimental softening by as much as $300$\% 
close to~$\Gamma$.
At a doping concentration of $1.4\cdot10^{21}~{\rm cm}^{-3}$, for instance, the adiabatic LO 
phonon energy at $\Gamma$ is softened by 
$\Delta \Omega^{\rm adiab}_\Gamma=22$~meV, whereas 
from IXS we have $\Delta \Omega^{\rm exp}_{{\Gamma}}=5.3$~meV. 
The non-adiabatic theory, on the other hand, 
yields a softening in excellent agreement with experiment: for instance, we obtain 
$\Delta \Omega^{\rm NA}_{{\Gamma}}=7$~meV for the same doping level. 
These results are further corroborated by considering 
an Einstein phonon model coupled to a homogeneous electron gas 
with parabolic dispersion $\epsilon_{\bf k}=\hbar^2k^2/2m^*_{\rm dos}$, with $m^*_{\rm dos}=1.18$  being the 
density-of-state effective mass of diamond. 
Within these approximations Eq.~(\ref{eq:PiNA}) reduces to 
$\hbar \Pi =2 g^2 \epsilon_\infty [ \chi_0 (\omega_{\rm E})- \chi_0 (0) ]$, 
with $\chi_0(\omega)$ being the long-wavelength limit (${\bf q}\rightarrow0$) 
of the Lindhard function \cite{Mahan2000}. 
For diamond, using $\hbar\omega_{\rm E}=0.16$~eV,  $g=0.1$~eV, $m^*_{\rm dos}=1.18$, and $\epsilon_\infty=5.44$, 
we obtain $\hbar\Pi\simeq8$~meV for $n=1.4\cdot10^{21}$~cm$^{-3}$,  in agreement with our ab initio calculations.

These features are also nicely reproduced by the phonon dispersions reported in 
Fig.~\ref{fig1}~(g)-(h), confirming the non-adiabatic character of the KA. 
Owing to the undamped nature of phonons in the adiabatic approximation 
(here we ignore phonon-phonon interactions), the adiabatic spectral 
functions are characterized by infinitesimal linewidths. 
The non-adiabatic spectra, on the other hand, correctly reproduce 
(i) the increase of spectral linewidth with doping concentration, and
(ii) the decrease of the linewidth with phonon momentum 
as shown in {Fig.~\ref{fig1}~(c)-(h) and in Fig.~S3 \cite{sup}}. 
The resulting spectral lineshapes are in good qualitative agreement with IXS, 
suggesting that electron-phonon scattering constitutes the primary mechanism 
for LO phonon damping in superconducting diamond.  

The pronounced non-adiabatic character of the lattice dynamics in doped diamond 
indicates a breakdown of the adiabatic Born-Oppenheimer approximation. 
This effect may be explained by considering the timescales involved: 
while LO phonons oscillate with a period $\tau_{\rm ph} = 25$~fs, the 
timescale of electronic screening $\tau_s$ is set by the plasma frequency $\omega_{\rm pl}$ via
$\tau_s = 2\pi/\omega_{\rm pl} =2\pi(4\pi n /m^*\epsilon_\infty)^{-1/2}$, 
with $m^*$ being the carrier effective mass. 
Using this expression, we find $\tau_s = 9$ and 4~fs for 
$n=3\cdot10^{20}$ and $1.4\cdot10^{21}~{\rm cm}^{-3}$, respectively, 
which are compatible with the results of optical measurements \cite{Bustarret2001,Ortolani2006}.
As screening operates on timescales that approach the characteristic 
phonon period, the assumptions underlying the Born-Oppenheimer approximation 
are not valid, and we see the emergence of strong non-adiabatic coupling. 

As a first step to explore the consequences of non-adiabaticity in B-doped diamond, 
we examine the superconducting critical temperature $T_{\rm c}$ using McMillan's formula 
\cite{McMillan,Allen1975}: 
$T_{\rm c} = {\langle \omega \rangle}/{1.2} \,{\rm exp} \lbrace
- 1.04 (1+\lambda)/[{\lambda-\mu^*(1 + 0.62\lambda)]}\rbrace$, 
where $\lambda$ is the electron-phonon coupling strength,
and $\langle \omega \rangle$ the logarithmic average of the phonon frequency.
{
Following Refs.~\cite{Scalapino1965,Allen1975},
the Coulomb pseudopotential $\mu^*$ is set to the standard value of $0.11$}.
Noting that $\lambda\propto\omega_{{\bf q}\nu}^{-2}$ 
\cite{giustino2016}, a small change in the 
phonon frequency as introduced by the adiabatic approximation, 
may induce a large modification of $T_{\rm c}$. 
At a doping concentration of $1.4\cdot10^{21}~{\rm cm}^{-3}$, for instance, 
the adiabatic approximation underestimates the LO phonon 
frequency in diamond by $\sim10\%$. In turn, this results
into an overestimation of $\lambda$ by $\sim20\%$. 
This inaccuracy is amplified by the exponential 
dependence of $T_{\rm c}$ on $\lambda$, leading to an overestimation of the 
critical temperature by up to $50\%$. 
Non-adiabatic effects thus carry important implications for the 
theoretical prediction of $T_{\rm c}$, and should be considered in future studies. 

In conclusion, by combining first-principles calculations of the electron-phonon interaction 
and high-resolution IXS experiments, we demonstrated the emergence of a 
non-adiabatic KA in superconducting diamond.
{ Beside resolving a long-standing discrepancy between theory
and experiment, these findings reveal that a breakdown of the Born-Oppenheimer 
approximation may lead to sizeable renormalization effects in the phonon 
dispersions of three-dimensional crystals. }
Our work calls for a systematic investigation of non-adiabatic effects
and Kohn anomalies in the phonon dispersions of three-dimensional 
heavily doped semiconductors as well as superconducting oxides.

\acknowledgements
We wish to thank L. Ort{\'e}ga for help with the x-ray diffraction characterisation 
of the samples and F. Jomard for calibration of the B-concentration by secondary 
ion mass spectrometry (SIMS) and depth profiling a few $\mu$m. 
The research leading to these results has received funding from the Leverhulme Trust (Grant RL-2012-001),
the European Union's Horizon 2020 research and innovation programme under grant agreement No.~696656 - GrapheneCore1, and
the UK Engineering and Physical Sciences Research Council
(Grant No. EP/J009857/1). Supercomputing time was provided by the University of Oxford Advanced Research
Computing facility (http://dx.doi.org/10.5281/zenodo.22558) and the ARCHER UK National Supercomputing Service.
We acknowledge the ESRF for granting use of beamline ID28, which contributed to the results presented here.


%

\end{document}